\documentclass[preprint,doublespacing,tightenlines,showpacs,showkeys,amsmath,amssymb,a4paper,12pt]{revtex4}

\usepackage{epsfig}
\usepackage{amssymb}
\usepackage{amssymb}
\usepackage{amsmath}
\usepackage{color}
\usepackage{subfigure}
\usepackage{psfrag}
\usepackage{cases}
\usepackage[retainorgcmds]{IEEEtrantools}

\begin{document}

\title{On the thermodynamics of volume/mass diffusion in fluids}

\author{S.\ Kokou Dadzie}
\email{k.dadzie@glyndwr.ac.uk}
\affiliation{Mechanical and Aeronautical Engineering, Glynd\^wr University, \\
Mold Road, Wrexham LL11 2AW, UK}
\author{Jason M.\ Reese}
\email{jason.reese@strath.ac.uk}
\affiliation{Department of Mechanical and Aerospace Engineering, University of Strathclyde, Glasgow G1 1XJ, UK}

\date{\today}

\begin{abstract}
In reference \cite{DadzieReese.PhysicaA.2008}, a kinetic equation for gas flows was proposed that leads to
a set of four macroscopic conservation equations, rather than the traditional set of three equations.
The additional equation arises due to local spatial random molecular behavior, which has been described as a volume or mass diffusion process.
In this present paper, we describe a procedure to construct a Gibbs-type equation and a second-law associated with these kinetic and continuum models.
We also point out the close link between the kinetic equation in \cite{DadzieReese.PhysicaA.2008} and that proposed previously by Klimontovich, and
we discuss some of their compatibilities with classical mechanical principles. Finally, a dimensional analysis highlights  the nature
of volume/mass diffusion: it is a non-conventional diffusive process, with some similarities to the `ghost effect', which cannot be
obtained from a fluid mechanical derivation that neglects non-local-equilibrium structures, as the conventional Navier-Stokes-Fourier model does.
\end{abstract}


\maketitle
\newpage

\section{Introduction}
In the continuity equations of fluid mechanics, as well as in the Boltzmann kinetic equation
for gases, spatial diffusive or source terms are not included.
This is widely accepted as representing local conservation of mass.
In fact, the continuity equation has a more general and purely
mathematical foundation, originating from conservation of a probability (actually, a probability density)
that is substituted for rationally-defined physical quantities in order to
derive a variety of well-known equations in physics.
In conventional kinetic theory (and also in fluid dynamics), the spatial molecular probability
density is taken to be equivalent to the physical mass-density of the fluid.
Consequently, the mathematical conservation of probability density, i.e. the continuity equation,
is taken to be the same as the evolution equation for the physical mass-density. Mass-density
at the macroscopic scale and in thermodynamics is taken to be an amount of mass divided by the volume containing that mass.

Questions regarding the role of microscopic molecular spatial fluctuations and molecular/continuum length scale differences have often
been raised by researchers \cite{OttingerHansJouStaPhy2010,BorisAlexeev,koga1954-Chem.Phy,Goddard2010_IntJouEngScie}.
Some went on to suggest modifications to both the Boltzmann kinetic equation and the continuum fluid equation set, including
dissipative components appearing in the mass-density equation \cite{BogDor2011,BorisAlexeev,KlimontovichPhyLet1992,Ueyama-JSP1980,koga1954-Chem.Phy,BardowOttingerPhysicaA-2007}.
However, these models are often accused of violating the fundamental concept of mass conservation, and so they usually do not attract much interest
within the fluid dynamics community.

In rarefied gas dynamics, there are flow systems where experimental evidence indicates that
existing gas flow models are not adequate, e.g. \cite{ShermanTalbotRGD1960,LengrandAllegreChpounRGD1996,TakuOhwada-PhyFlu1996,AgarwalYun2001,XJGu_DEmersonJFM2009,CaoSunChenGu02009}.
A number of these flow configurations (such as shock wave structure, sound wave propagation, the
mass-flowrate in a microchannel) are therefore useful testing-grounds for any newly-developed gas flow models. Sone
et al.\ \cite{SoneAokiTakataSugimoto_PhyFlu1996,SoneToshiyuki_PhyFlu2000,SoneToshiyuki_PhyFlu2004,SoneToshiyuki_PhyFlu2003} described a series of flow configurations that, although belonging
to the standard Navier-Stokes-Fourier regime, cannot be described by the conventional fluid model.
With increasing interest in micro/nano scale flows, experimental evidence from both liquids
and gases \cite{ewartJFM2007,MattiaGogotsiMicroNano2008} is extending the range of application domain in which the conventional fluid model fails.
There is, therefore, a substantial motivation to develop and test new flow models in order to understand the physics of flows beyond conventional fluid mechanics.

In the kinetic theory of gases, there are also recurrent problems faced by continuum models derived using approximation
solution techniques to the conventional Boltzmann kinetic equation to an order in Knudsen number beyond that of Navier-Stokes-Fourier.
These are \cite{StruchtrupTorrilhonPhysRevLett2010}: consistency with the second law of thermodynamics, frame indifference,
and angular momentum conservation. Woods extensively investigated these problems for Burnett-type equations \cite{LCWoodsBoltzmanLimitation2008,WoodJFM1983,LCWoods}.

In reference \cite{DadzieReese.PhysicaA.2008}, a kinetic model equation was proposed, following a claim of the existence
of a new form of diffusive process, so-called ``volume diffusion'' \cite{Brenner.PhysicaA.revs.2005,BrennerPhyDiffVol2010,Brenner_Phy_PPhysica_BiveloHydro2010}.
In this present paper, we show that it is possible to construct a Gibbs-type equation and a second-law associated with that kinetic model.
Then, using a dimensional analysis, we demonstrate that volume/mass diffusion terms in the fluid equations are non-conventional diffusive processes reflecting the microstructure behaviour beyond the accuracy order  of Navier-Stokes-Fourier.

\section{Stochastic kinetic equations and continuum-fluid volume diffusion}

\subsection{The Klimontovich spatial stochastic kinetic model}

The Boltzmann kinetic theory,  based on a single particle distribution function,
attaches no special importance to the spatial arrangements of molecules.
An average number of molecules is associated with a position variable $X$.
This position variable is then associated with the continuum fluid position variable.
No clear distinction exists between the kinetic-level spatial position and the hydrodynamic position variable.
Klimontovich questioned this aspect of the kinetic description, and introduced a separation scale in these two position variables \cite{KlimontovichPhyLet1992}. He assumed a microscopic distribution of the continuum volume given by a Gaussian.
Then, using some relaxation arguments and integration over the microscopic volume, arrived at a modified Boltzmann equation which differs from the original by a spatial diffusion term. Klimontovich's modified kinetic equation can be written:
\begin{equation}
\label{StochaKineticKokou}
\frac{\partial f}{\partial t} + \xi \cdot \nabla  f - \nabla \cdot \kappa \nabla f +  F_{ext} \cdot
\nabla_\xi f - I_{\xi}(f)  = 0 ,
\end{equation}
where $\kappa$ is a spatial diffusivity coefficient, $\xi$ is the kinetic particle drift velocity, $F_{ext}$ denotes an
external body force (not included originally in the Klimontovich expression),
 $\nabla_{\xi}$ is the gradient operator in velocity space, $I_{\xi}(f)$
is the Boltzmann collision integral, with $f(t,X,\xi)$ the (modified) distribution function of the molecules.
In fact, equation (\ref{StochaKineticKokou}) is a Fokker-Planck type of equation for diffusion with drifts,
where the diffusion is explicit in both position and velocity spaces, and the velocity space diffusion operator is substituted for
the Boltzmann collision integral.
The equation can also be derived using purely stochastic reasoning \cite{PKotelenez,RKuboJMP_1962,ChandrasekharRMP1943}.
On dropping the spatial diffusion term, it reverts to the classical deterministic Boltzmann equation.

A macroscopic continuum set of equations corresponding to kinetic equation (\ref{StochaKineticKokou}) can be derived by first
defining macroscopic flow variables as follows. The macroscopic mass-density $\rho(t,X) $ and macroscopic flow velocity $U(t,X)$ are
\begin{align}
\label{density_velocity_Klimon}
\rho(t,X) = \int M f (t, X, \xi) d_\xi,  \ \ \text{and} \  \ \rho(t,X)  U(t,X) = \int  M \xi f (t, X, \xi) d_\xi,
\end{align}
where $M$ is the molar mass.
Note that in equations (\ref{density_velocity_Klimon}), the integration is only over the microscopic molecular velocity $\xi$. The position variable $X$ is the same as that
appearing in kinetic equation (\ref{StochaKineticKokou}).
But, in the Klimontovich modification, variable $X$
in both equations (\ref{StochaKineticKokou}) and (\ref{density_velocity_Klimon}) represents a continuum position variable, not a position at the kinetic level of individual molecules.

The derivation of a continuum set of local transport equations follows by multiplying the stochastic kinetic equation (\ref{StochaKineticKokou})
by $M, M\xi, M\xi^2/2$, and integrating over velocity space only. This gives, respectively,
\begin{subequations}
\begin{align}
\label{massnewhydro} \frac{\partial  \rho}{\partial t} + \nabla \cdot [ \rho U  \underline{- \kappa  \nabla \rho }] =0 ,
\end{align}
\begin{align}
\label{momentumnewhydro} \frac{\partial  \rho U}{\partial t} + \nabla \cdot \left[
\rho U U\right] + \nabla \cdot \left[  p \mathbf{I}+ \Pi  \right] \underline{-  \nabla \cdot \left[\kappa \nabla \left(\rho U \right) \right]}= 0 ,
\end{align}
\begin{align}
\label{energyhydro}
\nonumber
 \frac{\partial }{\partial t}
\left[\frac{1}{2}  \rho U^2   +  \rho e_{in} \right]  +
 \nabla \cdot \left[\frac{1}{2} \rho U^2 U +  \rho
e_{in} U\right]
 + \nabla \cdot   \left[ \left(  p \mathbf{I} + \Pi \right)
\cdot U \right] + \nabla \cdot \left[\mathbf{q} \right] &  \\
\underline{-  \nabla \cdot \left[\kappa \nabla \left(\frac{1}{2}  \rho U^2   +  \rho e_{in} \right) \right]} & = 0  ,
\end{align}
\end{subequations}
where $\mathbf{I}$ is the identity matrix and $p$ is the kinetic pressure, which is related to internal energy by $3 p= 2 \rho e_{in}$.
The underlined terms in this set of equations are the new additions arising from the random change in molecular positions at the microscopic level, not from the random exchange of momentum. Quantities  $e_{in}$, $\Pi$ and $\mathbf{q}$ are all related to the random component of molecular velocity in the same way as in the traditional derivation from the deterministic Boltzmann equation. Essentially,
the shear stress $\Pi$ and heat flux $\mathbf{q}$ that result from molecular level diffusion of momentum and energy
can be given their Newton and Fourier law expressions. According to the kinetic theory definition, the kinetic temperature $T$
can also be associated  with $e_{in} = (3/2) R T$, with $R$  the specific gas constant.

The set of equations (\ref{massnewhydro}) to (\ref{energyhydro}) produces, for example, a better prediction of sound wave dispersion,
which is a flow configuration where microscopic structures are not as negligible as the Navier-Stokes-Fourier model requires \cite{DadzieReese.PhysLetter.2012}.
However, the same underlined terms above that generate acceptable flow results,
also appear apparently to contradict classical mechanic conservation principles and the second law of thermodynamics
(see Appendix \ref{AppendixA}).

Arguments for the modified kinetic equation (\ref{StochaKineticKokou}) focus on the introduction of the
spatial distributions of molecules. Assuming a Gaussian distribution of the continuum volume, as Klimontovich did,
avoids introducing an evolution equation for the continuum volume.
However, a microscopic variable could be introduced to describe the actual evolution of the molecular spatial distribution.
Such a full derivation is presented in \cite{DadzieReese.PhysicaA.2008}, and we now  analyze the thermodynamic consistency
of that volume model.

\section{Explicit volume-based kinetic and continuum equations}

\subsection{The kinetic and continuum equations, and mechanical consistency }

We follow a kinetic description with a distribution function $\mathfrak{f} (t, X,\xi, v )$
that describes the probability that a molecule at a
given time $t$ is located in the vicinity of position $X$, has its
velocity in the vicinity of $\xi$, and has around it a microscopic
empty space given by the additional variable $v$. The presence of the microscopic
variable $v$ is justified by the stochastic aspect in the microscopic picture of the spatial distribution of molecules.
It is chosen here as a microscopic volume in order to track the macroscopic volume occupied by the entire collection of molecules.

Then, admitting that the probability density $ \mathfrak{f} (t, X,\xi, v )$ must be conserved, we can derived a kinetic equation for $ \mathfrak{f}(t, X,\xi, v )$ as \cite{DadzieReese.PhysicaA.2008}:
\begin{equation}
\label{eq.kinetic.equation.kok}
\frac{\partial \mathfrak{f} }{\partial t} + (\xi \cdot \nabla ) \mathfrak{f} + W  \frac{\partial \mathfrak{f} }{\partial v} + F_{ext} \cdot
\nabla_\xi \mathfrak{f} - I_{\xi}(\mathfrak{f})  = 0 .
\end{equation}
Operator $I_{\xi}(\mathfrak{f})$, describing the molecular level diffusion of momentum,
is taken in equation (\ref{eq.kinetic.equation.kok})
as the Boltzmann hard sphere molecular collision integral;
this is because it operates only in velocity space, while $X$ and $\xi$ are independent in the same way as in kinetic equation (\ref{StochaKineticKokou}).
On the left-hand-side of equation (\ref{eq.kinetic.equation.kok})
the term containing $W$ now replaces the Laplacian spatial diffusion term in kinetic equation (\ref{StochaKineticKokou}).
This term embodies the evolution of the macroscopic continuum volume from the microscopic volumes,
and $W$ is the local continuum volume production rate. An
expression for $W$ needs to be defined, and it is a property belonging to a collection of molecules rather than an individual molecule.
The external force $F_{ext}$, which operates as a drift in velocity space and generates momentum and accelerates a molecular motion,
will be neglected for simplicity without changing any aspect of the fundamental description.

\subsubsection{The macroscopic field variables}
The introduction of an additional microscopic parameter modifies the method of constructing macroscopic field variables in our explicit volume-based kinetic description.

As $\mathfrak{f}(t,X,\xi,v)$ is presented primarily as a probability density function, we have a normalization factor (or a reduced probability density):
\begin{equation}
\label{Andefin}
A_n(t,X) =  \int  \int \mathfrak{f}(t, X, \xi, v
)  d_v d_\xi \ .
\end{equation}
The mean value, $\bar{Q}(t,X)$, of a gas property $Q$ is then
defined by,
\begin{equation}
\label{meanvalu}
\bar{Q}(t,X)   = \frac{1}{A_n(t,X)}  \int \int Q  \mathfrak{f}(t, X, \xi, v )
d_v d_\xi \ .
\end{equation}
We have then a local average of $v$ as the mean empty volume
$\bar{v}(t,X)$ around each gaseous molecule, i.e.
\begin{equation}
\label{meanvoldefi}  \bar{v} (t,X) = \frac{1}{A_n(t,X)}\int
\int v \mathfrak{f}(t, X, \xi, v )  d_v d_\xi \  .
\end{equation}

From equation (\ref{Andefin}) it appears that $A_n$ represents
a probable amount of mass (or, equivalently, number of molecules) around position $X$, but does not contain
 information about the distribution of that mass
around that position. How the amount of mass is spread out in space is given
through $\bar{v}(t,X)$ in equation (\ref{meanvoldefi}).
As average quantities defined via equation (\ref{meanvalu}) are mass-based averaged,
a definition of the physical mass-density of a fluid should be:
\begin{equation}
\label{massdensitydef}
\bar{\rho} (t,X)  =
\frac{M}{\bar{v}(t,X)}  \ ,
\end{equation}
where $M$ is the molecular mass.

Two mean velocities are defined
using two different weighting values: the local mean mass-velocity
$U_m (t,X)$ is given through:
\begin{equation}
\label{vitessemass}
A_n(t,X) U_m (t,X) = \int \int \xi \mathfrak{f}(t, X, \xi, v ) d_\xi d_v ,
\end{equation}
and a local mean volume-velocity $U_v(t,X)$ by using the
microscopic empty volume as the weighting, i.e.
\begin{equation}
\label{vitessevolume}
\bar{v} (t,X) A_n(t,X)  U_v (t,X) = \int \int  v \xi \mathfrak{f}(t, X, \xi, v ) d_\xi d_v .
\end{equation}
It is important to note that the two definitions for $U_v$ and $U_m$ coincide if $v$ is assumed constant.
For a homogeneous $v$ the present description is therefore equivalent to the
classical kinetic theory. For a non-homogeneous $v$,
we denote $U_v - U_m = \bar{v}^{-1}\mathbf{J}_v$ which evidently has dimensions of velocity. The quantity $U_v - U_m$ can physically represent either a volume or mass diffusion process \cite{DadzieReese.PhysicaA.2008}.

\subsubsection{Consistency with mechanics}
The separation of mass from volume in this kinetic construction introduces two types of
averaging that relate to the way balance equations are analysed in classical
continuum mechanics; namely, `control volume' and `control mass' concepts \cite{Lai_Krempl_Rubin}.
So the macroscopic set of derived continuum equations can be written in two versions.

By integrating the kinetic equation (\ref{eq.kinetic.equation.kok}) over $\xi$ and $v$, after multiplication by the different microscopic elements
$v , 1, \xi $ and $\xi^2$,
the set of macroscopic equations take the form:
\begin{subequations}
\begin{equation}
\label{volumediffuse}
\frac{\partial A_n \bar{v} }{\partial t} +  \nabla \cdot [A_n
\bar{v}U_m] +
 \nabla \cdot [A_n\mathbf{J}_v]  = A_nW,
\end{equation}
\begin{align}
\label{massnewmacro} \frac{\partial  A_n}{\partial t} + \nabla \cdot [ A_n U_m]  = 0,
\end{align}
\begin{align}
\label{momentumnewmacro} \frac{\partial  A_nU_m}{\partial t} + \nabla \cdot \left[
A_nU_mU_m\right] + \nabla \cdot [ A_n \mathfrak{P} ]  = 0,
\end{align}
\begin{align}
\label{energymacro}
 \frac{\partial }{\partial t}
\left[\frac{1}{2}  A_n U_m^2 \right]  + \frac{\partial}{\partial t}
\left[  A_n  \mathfrak{e_{in}}  \right]  +
 \nabla \cdot \left[\frac{1}{2} A_n U_m^2U_m +  A_n
\mathfrak{e_{in}} U_m\right] \\ \nonumber  + \nabla \cdot  \left[ A_n \mathfrak{P}
\cdot U_m\right] + \nabla \cdot [ A_n \mathfrak{q}]  = 0,
\end{align}
\end{subequations}
where $U_m$ is the average mass-velocity, and quantities $\mathbf{J}_v$, $\mathfrak{P}$, $\mathfrak{e_{in}}$ and $\mathfrak{q}$
are all quantities remaining after extracting their macroscopic parts due to $U_m$; so these quantities are all related to the random mass velocity $\xi-U_m$.
At this stage, the set of equations (\ref{massnewmacro}) to (\ref{energymacro})
has a strong conservative form related to mass; in fact, they are exactly the same as
the traditional expressions for local conservation of mass, momentum and energy. So there is
no violation of local angular momentum conservation,
or any other fundamental mechanical problem.
Separately, the actual volume of fluid containing the local  mass, momentum and energy
is described in a separate equation (\ref{volumediffuse}), which does not have
the conservation structure of the others because of the production term $W$.

The volume of a fluid is generally considered as a thermodynamic property, so its
variation should impact on other properties too.
The new flux of volume, $\bar{v}^{-1}\mathbf{J}_v$,
has the dimensions of velocity. Adding this velocity to the average mass motion $U_m$ in order to define convective fluxes,
we obtain a second form of the momentum and energy equations that accounts for the fluid volume variations.
The macroscopic set of equations (\ref{volumediffuse}) to (\ref{energymacro})
is therefore equivalent to:
\begin{subequations}
\begin{align}
\label{massnewhydrovol} \frac{D A_n }{ Dt} = -  A_n \nabla \cdot U_m  ,
\end{align}
\begin{align}
\label{densitynewhydrovol}
A_n  \frac{D \bar{v} }{ D t}
=- \nabla \cdot
[A_n \mathbf{J}_v] +  A_nW ,
\end{align}
\begin{align}
\label{momentumnewhydrovol} A_n \frac{D U_m }{ D t}  = -  \nabla \cdot A_n
\left(\mathbf{P'} -  \frac{1}{\bar{v}^2}\mathbf{J}_v \mathbf{J}_v
\right),
\end{align}
\begin{align}
\label{energyhydrovol}
 A_n \frac{D }{ D t} \left[ \frac{1}{2} U_m^2 + e'_{in} - \frac{1}{2\bar{v}^2}\mathbf{J}_v^2 \right]  = &
   - \nabla \cdot A_n \left[\mathbf{q'}
   + \frac{1}{\bar{v}} e'_{in} \mathbf{J}_v\ \right]
   \\ \nonumber
    &
   - \nabla
\cdot A_n \left[ \left(\mathbf{P'} - \frac{1}{\bar{v}^2}\mathbf{J}_v
\mathbf{J}_v \right)\cdot U_v\right],
\end{align}
\end{subequations}
where the material derivative is defined with respect to $U_m$ as $D/Dt =
\partial /\partial t + U_m \cdot~\nabla$, and $U_v = U_m +\bar{v}^{-1}\mathbf{J}_v $. In equations (\ref{massnewhydrovol}) to (\ref{energyhydrovol})
the fluid internal properties and fluxes accounting for not only macroscopic mass motion but also the variation of
its volume occupied become $\mathbf{P'}$ and $\mathbf{q'}$ for momentum and energy flux, and $e'_{in}$ for the internal energy.
It is also clear that the structure of the local transport equations for momentum and energy is now different.
Taking $\mathbf{P'}$ and $\mathbf{q'}$ as diffusive fluxes, they should follow the classical
phenomenological Fick's law, at least to a first order linear approximation. These are
derived later by considering compatibility with the second law.

Taking equation (\ref{momentumnewhydrovol}) to the cross product with a continuum position vector $X$ we write
\begin{align}
\label{momentumnewhydrovol-conserv1}
X \wedge A_n \frac{D U_m }{ D t}  = - X \wedge \nabla \cdot A_n
\left(\mathbf{P'} -  \frac{1}{\bar{v}^2}\mathbf{J}_v \mathbf{J}_v
\right),
\end{align}
which  is equivalent to
\begin{align}
\label{momentumnewhydrovol-conserv444}
A_n \frac{ D  }{ D t} \left[ X \wedge U_m \right]   = - X \wedge \nabla \cdot A_n
\left(\mathbf{P'} -  \frac{1}{\bar{v}^2}\mathbf{J}_v \mathbf{J}_v
\right).
\end{align}
Now, for a symmetrical second order tensor $\bar {\bar{\mathrm{T}}}$, the following property holds:
\begin{align}
\label{momentumnewhydrovol-conserv222}
X \wedge  \left[ \nabla \cdot \bar {\bar{\mathrm{T}}} \right] = \nabla \cdot \left[ X \wedge \bar {\bar{\mathrm{T}}} \right].
\end{align}
Pressure tensor $\mathbf{P'}$ is symmetric by definition (see its full expression in equation \ref{fluxes-new}).
Consequently, sum of tensors appearing on the right-hand-side of equation (\ref{momentumnewhydrovol-conserv444}) is symmetrical. Equation (\ref{momentumnewhydrovol-conserv444}) has therefore the following final form
\begin{align}
\label{momentumnewhydrovol-conserv333}
A_n \frac{ D }{ D t} \left[ X \wedge U_m \right]  = -  \nabla \cdot \left[ A_n X \wedge
\left(\mathbf{P'} -  \frac{1}{\bar{v}^2}\mathbf{J}_v \mathbf{J}_v \right) \right].
\end{align}
Equation (\ref{momentumnewhydrovol-conserv333}) shows that the volume-based momentum equation (\ref{momentumnewhydrovol}) still satisfy the principle of angular momentum conservation. Generally, because the velocity vector involved in continuity equation (\ref{massnewhydrovol}) is the same as that involved on the left-hand-side of momentum equation (\ref{momentumnewhydrovol}), the set of volumed-based equations (\ref{massnewhydrovol}) to (\ref{energyhydrovol}) satisfies following mechanical properties: galilean invariance - integrability - angular momentum - center-of-mass position; see explicit demonstrations in Appendix \ref{AppendixB}.

\subsection{Consistency with the second law \label{entropyconsitency}}

Evaluation of the second law of thermodynamics in this case requires an energy equation that accounts for volume variations.
First, the energy equation (\ref{energyhydrovol}) is transformed using the momentum equation (\ref{momentumnewhydrovol}) into:
\begin{align}
\label{energyhydro-rwc0}
  A_n \frac{D }{ D t} \left[ e'_{in}
 - \frac{1}{2\bar{v}^2}\mathbf{J}_v^2 \right]   + \nabla \cdot A_n \left[ \left( e'_{in} -
  \frac{1}{\bar{v}^2}\mathbf{J}_v^2\right) \frac{1}{\bar{v}} \mathbf{J}_v \right] = \\ \nonumber
  - \nabla \cdot A_n \left[\mathbf{q'}
+\mathbf{P'} \cdot \frac{1}{\bar{v}} \mathbf{J}_v\right]
 - A_n\left(\mathbf{P'} -  \frac{1}{\bar{v}^2}\mathbf{J}_v \mathbf{J}_v
\right) : \nabla  U_m
 \ .
\end{align}
Then, using the continuity equation (\ref{massnewhydrovol}) and $\bar{\rho} \mathbf{P'} = p' \mathbf{I}+ \mathbf{\Pi_v}$, with $p'$
the kinetic theory definition of pressure by, $3 p'= 2 \bar{\rho} e_{in}'$, we obtain
\begin{align}
\label{energyhydro-rwc1}
  A_n \frac{D }{ D t} \left[ e'_{in}
 - \frac{1}{2\bar{v}^2}\mathbf{J}_v^2 \right]  - \frac{p'}{\bar{\rho}} \frac{D  A_n}{D t}  + \nabla \cdot A_n \left[ \left( e'_{in} -
  \frac{1}{\bar{v}^2}\mathbf{J}_v^2\right) \frac{1}{\bar{v}} \mathbf{J}_v \right] = \\ \nonumber
  - \nabla \cdot A_n \left[\mathbf{q'}
+\mathbf{P'} \cdot \frac{1}{\bar{v}} \mathbf{J}_v\right]
 - A_n\left( \frac{1}{\bar{\rho}} \mathbf{\Pi_v} -  \frac{1}{\bar{v}^2}\mathbf{J}_v \mathbf{J}_v
\right) : \nabla  U_m
 \ .
\end{align}
The volume evolution equation (\ref{densitynewhydrovol}) may be rewritten in terms of the density defined in
equation (\ref{massdensitydef}) as,

\begin{eqnarray}
\label{densitynewhydrovol-dens2}
 \frac{1}{\bar{\rho} }\frac{D A_n  }{ D t}
= \frac{D A_n \bar{\rho}^{-1} }{ D t} + \frac{1}{M}\nabla \cdot
[A_n \mathbf{J}_v] - \frac{A_n}{M} W  ,
\end{eqnarray}
which, substituted in equation (\ref{energyhydro-rwc1}), leads to:
\begin{align}
\label{energyhydro-rwc2}
  A_n \frac{D }{ D t} \left[ e'_{in}
 - \frac{1}{2\bar{v}^2}\mathbf{J}_v^2 \right]  - p'
 \frac{D A_n \bar{\rho}^{-1} }{ D t}  =   &  \frac{p'}{M}\nabla \cdot [A_n \mathbf{J}_v] - p' \frac{A_n}{M} W \\  \nonumber
                                          & - \nabla \cdot A_n \left[ \left( e'_{in}
                                          - \frac{1}{\bar{v}^2}\mathbf{J}_v^2\right) \frac{1}{\bar{v}} \mathbf{J}_v \right] \\ \nonumber
                                        &  - \nabla \cdot A_n \left[\mathbf{q'}
                                        +\mathbf{P'} \cdot \frac{1}{\bar{v}} \mathbf{J}_v\right] \\  \nonumber
                                        & - A_n\left( \frac{1}{\bar{\rho}} \mathbf{\Pi_v} -  \frac{1}{\bar{v}^2}\mathbf{J}_v \mathbf{J}_v
                                    \right) : \nabla  U_m  \ .
\end{align}
This can be rearranged using $e'_{in} =3/2 (p'/\bar{{\rho}})$ into :
\begin{align}
\label{entropy-volume0}
  A_n \frac{D }{ D t} \left[ e'_{in}
 - \frac{1}{2\bar{v}^2}\mathbf{J}_v^2 \right]  - p'
 \frac{D A_n \bar{\rho}^{-1} }{ D t}
                     = & - p' \frac{A_n}{M} W  - \nabla \cdot \left[\frac{A_n}{\bar{\rho}\bar{v}}\mathbf{\Pi_v} \cdot  \mathbf{J}_v\right]
                    + \frac{A_n}{\bar{v}^2}\mathbf{J}_v \mathbf{J}_v : \nabla  U_m \\ \nonumber
                      & +\frac{p'}{M}\nabla \cdot
                        [A_n \mathbf{J}_v] + \nabla \cdot \left[\left( - \frac{A_n p'}{\bar{\rho}}
                        +  \frac{A_n}{\bar{v}^2}\mathbf{J}_v^2 \right) \frac{1}{\bar{v}} \mathbf{J}_v \right] \\ \nonumber
                    &- \nabla \cdot \left[ A_n \mathbf{q'} +  \frac{3}{2}\frac{A_n p'}{\bar{\rho}}  \frac{1}{\bar{v}} \mathbf{J}_v \right] \\ \nonumber
                    & -  \frac{A_n}{\bar{\rho}} \mathbf{\Pi}_{Um} : \nabla  U_m
                    -  \frac{A_n}{\bar{\rho}} \mathbf{\Pi_{Jv}} : \nabla  U_m  \ ,
\end{align}
where the final three terms arise because the shear stress has been split into its components due to $U_m$ and due to $\mathbf{J}_v$,
respectively, $\mathbf{\Pi}_{Um}$ and $\mathbf{\Pi_{Jv}}$, such that $\mathbf{\Pi_v} = \mathbf{\Pi}_{Um}+\mathbf{\Pi_{Jv}}$, as  $U_v=U_m +\bar{v}^{-1} \mathbf{J}_v$ .

From the structure of the energy equation (\ref{entropy-volume0}),
we define an entropy quantity $\bar{s}$ such that:
\begin{align}
\label{Gibbsmodified}
  A_n \frac{D }{ D t} \left[ e'_{in}
 - \frac{1}{2\bar{v}^2}\mathbf{J}_v^2 \right]  - p'
 \frac{D A_n \bar{\rho}^{-1} }{ D t}
 =  A_n T' \frac{D \bar{s} }{ D t},
 \end{align}
 with $T'$ the associated temperature. The entropy equation obtained after substituting this
 definition into the energy equation (\ref{entropy-volume0}) is,
 \begin{align}
\label{entropy-volume1}
  A_n T' \frac{D \bar{s} }{ D t}
 =            - p' \frac{A_n}{M} W &
            - \nabla \cdot \left[\frac{A_n}{\bar{\rho}\bar{v}}\mathbf{\Pi_v} \cdot  \mathbf{J}_v\right]
            + \left( \frac{A_n}{\bar{v}^2}\mathbf{J}_v \mathbf{J}_v
            -  \frac{A_n}{\bar{\rho}} \mathbf{\Pi_{Jv}} \right) : \nabla  U_m
            \\ \nonumber
                   & +\frac{p'}{M}\nabla \cdot
            [A_n \mathbf{J}_v] + \nabla \cdot \left[\left( - \frac{A_n p'}{\bar{\rho}}
            +  \frac{A_n}{\bar{v}^2}\mathbf{J}_v^2 \right) \frac{1}{\bar{v}} \mathbf{J}_v \right]
            \\ \nonumber
                   & - \nabla \cdot \left[ A_n \mathbf{q'} +  \frac{3}{2}\frac{A_n p'}{\bar{\rho}}  \frac{1}{\bar{v}} \mathbf{J}_v \right]
             -  \frac{A_n}{\bar{\rho}} \mathbf{\Pi}_{Um} : \nabla  U_m
            \ .
\end{align}
We have here a non-local-equilibrium entropy defined via equation (\ref{Gibbsmodified}), which is a Gibbs-like equation,
and an entropy evolution equation (\ref{entropy-volume1}) in which
the volume production rate $W$ is assumed to be a function of the macroscopic thermodynamic properties,
but with (so far) an unknown constitutive equation.
It follows that the first and second laws of thermodynamics can be used to construct an equation for $W$ from equation
(\ref{entropy-volume1}).

First, note that if we set
\begin{align}
\label{volumeProEqu}
 p' \frac{A_n}{M} W
 = &
 - \nabla \cdot \left[\frac{A_n}{\bar{\rho}\bar{v}}\mathbf{\Pi_v} \cdot  \mathbf{J}_v\right]
+ \left( \frac{A_n}{\bar{v}^2}\mathbf{J}_v \mathbf{J}_v \right) : \nabla  U_m
+  \frac{A_n}{\bar{\rho}} \mathbf{\Pi_{Jv}} : \nabla  \left[\frac{1}{\bar{v}}\mathbf{J}_v\right]
 \\ \nonumber
&  +  \frac{A_n}{\bar{\rho}} \mathbf{\Pi}_{Um} : \nabla  \left[\frac{1}{\bar{v}}\mathbf{J}_v\right]
 +\frac{p'}{M}\nabla \cdot
[A_n \mathbf{J}_v] + \nabla \cdot \left[\left( - \frac{A_n p'}{\bar{\rho}}
+  \frac{A_n}{\bar{v}^2}\mathbf{J}_v^2 \right) \frac{1}{\bar{v}} \mathbf{J}_v \right]
,
\end{align}
the entropy equation (\ref{entropy-volume1}) becomes,
\begin{align}
\label{entropy-volumesec2}
  A_n T' \frac{D \bar{s} }{ D t}
 =  - \nabla \cdot \left[ A_n \mathbf{q'} +  \frac{3}{2}\frac{A_n p'}{\bar{\rho}}  \frac{1}{\bar{v}} \mathbf{J}_v \right]
 -  \frac{A_n}{\bar{\rho}} \mathbf{\Pi}_{Um} : \nabla  U_m  -  \frac{A_n}{\bar{\rho}} \mathbf{\Pi_{Jv}} : \nabla  \left[\frac{1}{\bar{v}}\mathbf{J}_v\right]
 \\ \nonumber
 -  \frac{A_n}{\bar{\rho}} \mathbf{\Pi}_{Um} : \nabla  \left[\frac{1}{\bar{v}}\mathbf{J}_v\right]
 -   \frac{A_n}{\bar{\rho}} \mathbf{\Pi_{Jv}} : \nabla  U_m ,
\end{align}
or equivalently,
\begin{align}
\label{entropy-volumesec3}
 A_n T' \frac{D \bar{s} }{ D t}
 =  - \nabla \cdot \left[ A_n \mathbf{q'} +  \frac{3}{2}\frac{A_n p'}{\bar{\rho}}  \frac{1}{\bar{v}} \mathbf{J}_v \right]
 -  \frac{A_n}{\bar{\rho}} \mathbf{\Pi_v} : \nabla  U_v .
\end{align}
Equation (\ref{entropy-volumesec3}) has a structure such that entropic heat flux $\mathbf{q_s} $  is given by,
  \begin{align}
\label{entropy-heat-flux}
A_n \mathbf{q_s} = \left[ A_n \mathbf{q'} +  \frac{3}{2}\frac{A_n p'}{\bar{\rho}}  \frac{1}{\bar{v}} \mathbf{J}_v \right].
\end{align}
According to Linear Irreversible Thermodynamics concepts, and following equation (\ref{entropy-volumesec3}),
 the first order linear representation of the diffusive shear stress and entropic heat flux can be given according to a Fick's Law type of representation. These are then written,
\begin{subequations}
\begin{align}
\label{fluxes-new} \mathbf{\Pi_v} & = -  \mu'
\left( \frac{\partial U_{v_i}}{\partial X_j} + \frac{\partial
U_{v_j}}{\partial X_i}\right) + \eta' \frac{\partial
U_{v_k}}{\partial X_k}\delta_{ij} \ ,
\end{align}
\begin{align}
  \label{fluxes-new-mid}
\mathbf{q'}  & =  - \frac{\kappa'_h}{\bar{\rho}}  \nabla T'  - \frac{3}{2}\frac{p'}{\bar{\rho}}  \frac{1}{\bar{v}} \mathbf{J}_v  \ ,
\end{align}
\begin{align}
 \frac{1}{\bar{v}}\mathbf{J}_v & = - \frac{\kappa_m}{\bar{v}}   \nabla \bar{v}
 = \frac{\kappa_m}
{\bar{\rho}} \nabla \bar{\rho}\ ,
 \label{fluxes-new-end}
\end{align}
\end{subequations}
where $U_v = U_m +\bar{v}^{-1}\mathbf{J}_v $. The coefficients $\mu'$, $\kappa'_h$, $\eta'$ and  $\kappa_m$ are, respectively, dynamic viscosity, heat conductivity, bulk viscosity, and the mass-density diffusion coefficient in this volume-based description. As the kinetic pressure $p'$ was defined by the trace of the pressure tensor, we also have $(2/3)\mu' - \eta'=0 $ assuring the symmetric-traceless property of the shear stress tensor.
The entropy evolution equation (\ref{entropy-volumesec3}) may now be rewritten:
\begin{align}
\label{entropy-volumesec4}
 A_n T' \frac{D \bar{s} }{ D t}
 =  \nabla \cdot \left[ \frac{A_n}{\bar{\rho}}  \kappa'_h  \nabla T' \right]
 -  \frac{A_n}{\bar{\rho}} \mathbf{\Pi_v} : \nabla  U_v.
\end{align}
In summary, with an entropy evolution equation (\ref{entropy-volumesec4})
and the volume production term given by equation (\ref{volumeProEqu}), we have ensured that the set of volume-based continuum equations is thermodynamically consistent with the first order linear expressions for shear stress and heat flux. Indeed, the structure of the entropic heat flux in equation (\ref{entropy-heat-flux}) is such that the transport coefficients satisfy the Onsager reciprocal relations, and the entropy production rate follows a bilinear structure. Then
non-negativity of the coefficients $\mu'A_n/\bar{\rho}$ and $\kappa'_hA_n/\bar{\rho}$ assures the non-negativity of the entropy production with a shear stress expressed as in equation (\ref{fluxes-new}). However, we note that, to satisfy these properties, constructing $W$  as it is given in equation (\ref{volumeProEqu}) is not a unique option.

The expression for the volume production, equation (\ref{volumeProEqu}), can now
be substituted into the volume equation (\ref{densitynewhydrovol})
to finalize the evolution equation for the specific volume of the fluid:
\begin{align}
\label{densitynewhydrovolfini}
A_n  \frac{p'}{M}\frac{D \bar{v} }{ D t}
=
        - \nabla \cdot \left[\frac{A_n}{\bar{\rho}\bar{v}}\mathbf{\Pi_v} \cdot  \mathbf{J}_v\right]
        & + \left( \frac{A_n}{\bar{v}^2}\mathbf{J}_v \mathbf{J}_v \right) : \nabla  U_m
                +  \frac{A_n}{\bar{\rho}} \mathbf{\Pi_v} : \nabla  \left[\frac{1}{\bar{v}}\mathbf{J}_v\right]
                \\ \nonumber
        & + \nabla \cdot \left[\left( -\frac{A_n p'}{\bar{\rho}}
        +  \frac{A_n}{\bar{v}^2}\mathbf{J}_v^2 \right) \frac{1}{\bar{v}} \mathbf{J}_v \right]
,
\end{align}
It is easily seen in this equation that a vanishing volume flux also means a constant $\bar{v}$ in the material reference frame. In such circumstances, all additional terms due to volume evolution in the momentum equation (\ref{momentumnewhydrovol}), as well as in constitutive equations (\ref{fluxes-new}) to (\ref{fluxes-new-end}), cancel out. The set of volume-based equations then simply returns
to the traditional Navier-Stokes-Fourier set, with a mass-density represented by $MA_n$.

\section{\label{sec_rotating}{The nature of the new diffusive volume/mass terms}}

\subsection{\label{sec_rotating}{The problem of a fluid in rotating equilibrium  }}

A configuration which was used to point out the violation of mechanical properties 
by volume/mass-diffusion hydrodynamic models is given in \cite{LiuMarioPhysRevLett2008ComOttin}.
The configuration is a fluid or gas in rotating equilibrium, at
constant temperature. More precisely, for a fluid rotating in equilibrium at constant temperature, contributions from $\nabla T'$ and the symmetric-traceless Navier-Stokes shear stress component related to $\nabla U_m$ vanish. Entropy equation (\ref{entropy-volumesec4}), considered without any preliminary analysis of the nature of the terms involved, can then be reduced to:
\begin{align}
\label{entropy-volumesec-violation}
 A_n T' \frac{D \bar{s} }{ D t}
 =  -  \frac{A_n}{\bar{\rho}} \mathbf{\Pi_{Jv}} : \nabla  \left[\frac{1}{\bar{v}}\mathbf{J}_v\right].
\end{align}
Equation (\ref{entropy-volumesec-violation}) shows a production of entropy for an apparently equilibrium configuration. This is, according to \cite{LiuMarioPhysRevLett2008ComOttin}, not acceptable.

In fact, whenever additional local microscopic structures are involved, additional entropy production in the way it appears in equation
(\ref{entropy-volumesec-violation}) should be expected (see existing methods for non-local-equilibrium fluids, for example \cite{ottingerbook,Vilensky2011,HutterBrader2009_TheJouChePhy,TKoide_TKodama_arxiv2011}).
We now demonstrate using a dimensional analysis that the production term in equation (\ref{entropy-volumesec-violation}) is
indeed a type of non-local-equilibrium term which is negligible at the order of approximation of the Navier-Stokes-Fourier model of hydrodynamics.

\subsection{A dimensional analysis of the new entropy (energy) equation}
Denoting the gas mean free path $\lambda$, we define the following reference fluid parameters:
\begin{align}\label{reference}
& \mu^0 = \rho^0  C^0 \lambda, \quad \kappa_m^0 = \frac{\mu^0}{\rho^0}, \quad t^0= \frac{\lambda}{C^0} , \\
& \kappa_h^0 = \frac{\mu^0 {C^0}^2} { T^0}, \quad \rho^0 = MA_n^0, \notag
\end{align}
where superscript `$^0$' refers to a reference characteristic flow parameter, and $C^0$ is a characteristic molecular agitation speed.
Then we define dimensionless flow properties through:
\begin{align}\label{reference-dimless}
& \mu=\mu^*\mu^0,  \quad \kappa_m = {\kappa_m}^* {\kappa_m}^0, \quad \kappa'_h = {\kappa_h}^* \kappa_h^0, \quad  A_n=A_n^*A_n^0,  \\
& \rho = \rho^* \rho^0, \quad p =p^* p^0, \quad T = T^* T^0, \quad x=x^*L, \notag
\end{align}
where superscript `$ ^* $' indicates a dimensionless quantity, and $L$ a macroscopic length scale. Entropy equation (\ref{entropy-volumesec4}) can then be rewritten
\begin{align}
\label{entropy-volumesec4-dimless}
\frac{MA_n^0 T^0 s^0}{t^0} A_n^* T^* \frac{D \bar{s}^* }{ D t^*}
 = &  \frac{\kappa_h^0 T^0 }{L^2} \nabla \cdot \left[ \frac{A_n^*}{\bar{\rho}^*}  \kappa_h^*  \nabla T^* \right]
    -  \frac{\mu^0 {C^0}^2}{L^2} \frac{A_n^*}{\bar{\rho^*}} \mathbf{\Pi}_{Um}^* : \nabla  U_m^*
        \\ \nonumber
    &  -  \frac{{\mu^0}^2 C^0} {\rho^0 L^3} \frac{A_n^*}{\bar{\rho^*}} \mathbf{\Pi}_{Um}^* : \nabla  \left[ \frac{\kappa_m^*} {\bar{\rho}^*} \nabla \bar{\rho}^* \right]
        -   \frac{{\mu^0}^2 C^0} {\rho^0 L^3} \frac{A_n^*}{\bar{\rho^*}} \mathbf{\Pi^*_{Jv}} : \nabla  U_m^*
    \\ \nonumber
    & -  \frac{{\mu^0}^3}{{\rho^0}^2 L^4} \frac{A_n^*}{\bar{\rho^*}} \mathbf{\Pi^*_{Jv}}
        : \nabla  \left[ \frac{\kappa_m^*}
        {\bar{\rho}^*} \nabla \bar{\rho}^* \right] ,
\end{align}
which can be re-arranged using the expression for $\mu^0$ and $\kappa_h^0$ from equation (\ref{reference}) as
\begin{align}
\label{entropy-volumesec4-dimless2}
\frac {L^2}{\mu^0 {C^0}^2} \frac{MA_n^0 T^0 s^0}{t^0} A_n^* T^* \frac{D \bar{s}^* }{ D t^*}
  =& \nabla \cdot \left[ \frac{A_n^*}{\bar{\rho}^*}  \kappa_h^*  \nabla T^* \right]
 - \frac{A_n^*}{\bar{\rho^*}} \mathbf{\Pi}_{Um}^* : \nabla  U_m^*
 \\ \nonumber
 & -  K_n \frac{A_n^*}{\bar{\rho^*}} \mathbf{\Pi}_{Um}^* : \nabla  \left[ \frac{\kappa_m^*} {\bar{\rho}^*} \nabla \bar{\rho}^* \right]
 -  K_n \frac{A_n^*}{\bar{\rho^*}} \mathbf{\Pi^*_{Jv}} : \nabla  U_m^*
 \\ \nonumber
  &   -  K_n^2 \frac{A_n^*}{\bar{\rho^*}}
     \mathbf{\Pi^*_{Jv}} : \nabla  \left[ \frac{\kappa_m^*}
    {\bar{\rho}^*} \nabla \bar{\rho}^* \right],
 \end{align}
in which appears the Knudsen number, $ K_n= \lambda / L $. Defining the reference entropy per unit-mass $s^0$, in a way that the unit-volume of fluid is re-scaled by the mean free volume $\lambda^3$, i.e.,
 \begin{align}\label{entropy-zero}
 \frac{s^0}{\lambda^3}=  \frac{1}{LT^0 {t^0}^2 },
 \end{align}
entropy equation (\ref{entropy-volumesec4-dimless2}) then yields the following full dimensionless form:
\begin{align}
\label{entropy-volumesec4-dimless3}
A_n^* T^* \frac{D \bar{s}^* }{ D t^*}
 = &  K_n \nabla \cdot \left[ \frac{A_n^*}{\bar{\rho}^*}  \kappa_h^*  \nabla T^* \right]
        - K_n \frac{A_n^*}{\bar{\rho^*}} \mathbf{\Pi}_{Um}^* : \nabla  U_m^*
        \\ \nonumber
    & -  K_n^2 \frac{A_n^*}{\bar{\rho^*}} \mathbf{\Pi}_{Um}^* : \nabla  \left[ \frac{\kappa_m^*} {\bar{\rho}^*} \nabla \bar{\rho}^* \right]
        -  K_n^2 \frac{A_n^*}{\bar{\rho^*}} \mathbf{\Pi^*_{Jv}} : \nabla  U_m^*
            \\ \nonumber
    &   -  K_n^3 \frac{A_n^*}{\bar{\rho^*}} \mathbf{\Pi^*_{Jv}} : \nabla  \left[ \frac{\kappa_m^*}
            {\bar{\rho}^*} \nabla \bar{\rho}^* \right].
 \end{align}
In equation (\ref{entropy-volumesec4-dimless3}) it appears that all volume/mass diffusion terms are of higher order than Navier-Stokes-Fourier. The term related to the entropy production observed in the rotating fluid problem in equation (\ref{entropy-volumesec-violation}), is of order $K_n^3$.

In kinetic theory the Knudsen number is the parameter that quantifies the degree of non-equilibrium. It is considered as a small parameter when deriving continuum hydrodynamic models, as in the Chapman-Enskog expansion technique. So truncating equation (\ref{entropy-volumesec4-dimless3}) at order $K_n^0$, we get the zero entropy production that is consistent with the Euler hydrodynamic regime. Truncated at order $K_n$, equation (\ref{entropy-volumesec4-dimless3}) becomes exactly
the Navier-Stokes-Fourier entropy evolution equation. All volume/mass diffusion terms appear as non-equilibrium effects at an approximation level higher than that of  Navier-Stokes-Fourier.

Sone  et al.\ \cite{SoneToshiyuki_PhyFlu2004,SoneToshiyuki_PhyFlu2003,SoneAokiTakataSugimoto_PhyFlu1996,SoneToshiyuki_PhyFlu2000}
have described a series of flows that, while belonging to the continuum fluid regime, the conventional fluid mechanical model is incapable of predicting. Instead, one has to rely on additional higher order Knudsen number terms -- traditionally called rarefaction regime terms -- in order to predict the flow behaviour. The dimensional analysis above presents a clear exposition of these intriguing situations. Indeed, in the derivation of the volume/mass diffusion model we did not place any restriction or assumption on the Knudsen number. So, there may be configurations where these high order terms may dominate the lower order terms. This may be the case for flows with $K_n \geq 1$. In other words, our volume/mass diffusion terms are not of the same type as the conventional macroscopic Navier-Stokes-Fourier diffusion terms; they are effects arising from another level of microscopic behavior that may affect the global flow structure (or may not) depending on the configuration. In the case of the rotating fluid in equilibrium, they do not. Derivations of fluid mechanical models that neglect local microscopic structures, as many do cannot capture these types of physical effects as they simply do not account for the material structure at that level.

\section{The volume/mass diffusion model in an apparent continuum limit regime}

From our analysis of the nature of diffusive volume/mass terms, it appears that to account for
only the first level of this non-local-equilibrium embodied in the new diffusive volume process, we should
naturally neglect terms of high order $\mathbf{J}_v$, i.e. assume terms such as $\mathbf{J}_v \mathbf{J}_v $ to be negligible.
In so doing, we may also argue that at this first level, while a detailed evolution equation of the new local microscopic volume
may not be important by itself, it
still effects the macroscopic flow, mass, momentum and energy equations. In such circumstances, equations (\ref{massnewhydrovol}) to (\ref{energyhydrovol}) may be reduced to a three set of equations written:
\begin{subequations}
\begin{align}
\label{massnewhydrovolBren} \frac{D \bar{\rho} }{ Dt} = - \bar{\rho} \nabla \cdot U_m \ ,
\end{align}
\begin{align}
\label{momentumnewhydrovolbingo}
\bar{\rho}\frac{D U_m }{ D t}  =  - \nabla p' - \nabla \cdot  \mathbf{\Pi_v},
\end{align}
\begin{align}
\label{energyhydrovolbingo}
 \bar{\rho} \frac{D }{ D t} \left[ \frac{1}{2} U_m^2 + e'_{in}  \right] & =  - \nabla
\cdot   \left[ p'U_v + \mathbf{\Pi_v} \cdot U_v\right]
   - \nabla \cdot \left[\bar{\rho} \mathbf{  q'}
     +\frac{3}{2} p' \frac{1}{\bar{v}} \mathbf{J}_v\ \right],
\end{align}
\end{subequations}
closed with
\begin{subequations}
\begin{align}
\label{fluxes-bingo-stress}
\mathbf{\Pi_v}_{ij} & =  -  \mu'
\left( \frac{\partial U_{v_i}}{\partial X_j} + \frac{\partial
U_{v_j}}{\partial X_i}\right) + \eta' \frac{\partial
U_{v_k}}{\partial X_k}\delta_{ij} \ ,
 \\
\label{fluxes-bingo-heat}
\mathbf{q'}  & =  - \frac{\kappa'_h }{\bar{\rho}} \nabla T'   - \frac{3}{2} \frac{p'}{\bar{\rho}}  \frac{1}{\bar{v}} \mathbf{J}_v \\
 U_v &  = U_m + \frac{1}{\bar{v}}\mathbf{J}_v  = U_m + \frac{\kappa_m}
{\bar{\rho}} \nabla \bar{\rho},
\label{fluxes-bingo-volume}
\end{align}
\end{subequations}
where $MA_n = \bar{{\rho}}=M / \bar{v}$ is identified with the fluid physical local thermodynamic density and the evolution equation for the microscopic volume disregarded. This set of equations is exactly the same as proposed more recently also by Brenner
\cite{Brenner_Diffusevolutrans2010,Brenner_Phy_PPhysica_BiveloHydro2010}, using entirely classical irreversible thermodynamics reasoning.
The set of equations (\ref{massnewhydrovolBren})-(\ref{fluxes-bingo-volume}), as opposed to other previous attempts of volume or mass diffusion models \cite{OttingerHennigMarioPhysRevE2009,BardowOttingerPhysicaA-2007,ChakrabortyDurst-PhyFluids-2007,DadzieReese.PhysLetter.2012}
 can be said fully mechanically consistent (see explicit demonstrations in Appendix \ref{AppendixB}). Note that, mass conservation equation (\ref{massnewhydrovolBren}) does not involve explicitly a diffusive mass flux component. Instead, the volume/mass diffusion appears via the energy equation and in the expression of the shear stress tensor. If the volume velocity from equation (\ref{fluxes-bingo-volume})  is substituted for the mass velocity in equation (\ref{massnewhydrovolBren}),
 then we observe evidently a form of diffusive component in the mass equation. However, we note that this is not a true physical `dissipation' or `disappearance' of the mass as that diffusion is regarded with respect to the volume velocity rather than the mass velocity. Our volume-based momentum equation (\ref{momentumnewhydrovolbingo}), closed with equation (\ref{fluxes-bingo-stress}), is also the same as momentum equation derived recently in \cite{TKoide_TKodama_arxiv2011}. In that paper, the authors  presented the derivation using a different explicit spatial stochastic approach but did not provide the complete accompanying form of their energy or entropy evolution equations.

Finally, the two kinetic model equations (\ref{StochaKineticKokou}) and (\ref{eq.kinetic.equation.kok})
introduce into the kinetic description a stochasticity in position that does not appear, at least explicitly,
in the Boltzmann kinetic approach. Equation (\ref{StochaKineticKokou}) incorporates  this only as a diffusive volume, while
equation (\ref{eq.kinetic.equation.kok}) incorporates it in a more complete picture
as a convected volume and a volume source term.
Consequently, equation (\ref{StochaKineticKokou}) may be regarded as a reduced form of equation (\ref{eq.kinetic.equation.kok})
and both kinetic equations relate to a description in which local-equilibrium does not necessarily hold.

Without the volume source term in equation (\ref{entropy-volume1}), it would be challenging to
construct a second law analysis. As a matter of fact, constructing a second law for high order hydrodynamics beyond that of Navier-Stokes-Fourier while maintaining their mechanical consistency (e.g., for the Burnett equations) has always been a challenge \cite{ComeauxChapmanMacCormackAIAA95-0415,WoodJFM1983,LCWoods,StruchtrupTorrilhonPhysRevLett2010}. It appears that this new theory of volume diffusion and the   distinction between different types of velocities allow to achieve this. The procedure described here to obtain a version of the second law for this volume-based approach (in which  the additional volume/mass terms are of Burnett order), therefore opens
the door to reinvestigating longstanding problems with most, if not all, extended hydrodynamic models.

\section{Conclusion}
In this paper we have successfully constructed a thermodynamic second law for a volume-based hydrodynamic model
presented first in \cite{DadzieReese.PhysicaA.2008}. We have also provided further analysis of
the nature of the volume/mass diffusion process that is still currently a subject of active debate.
We have demonstrated that this diffusive process is not a conventional diffusion of the Navier-Stokes-Fourier type;
it arises from a deeper level of microscopic behaviour that may or may not affect the global flow structure, depending on the flow configuration.
Generally, a derivation that neglects local microstructure, as most conventional fluid mechanical derivations do, cannot capture this.

Future work may include:
\begin{itemize}
  \item reinvestigating thermodynamic problems related to Burnett and other extended hydrodynamic equations;
  \item comparing and contrasting the various volume/mass diffusion models proposed so far;
  \item investigating `ghost effect' flow configurations, which are flows in the pure continuum regime that require high order Knudsen number terms to be predicted appropriately -- these are therefore an ideal testing ground for volume/mass hydrodynamic models.
\end{itemize}

\section{Appendix: \label{AppendixA} Thermodynamic and mechanical consistencies of the Klimontovich volume model}

Let us assume a `material' derivative as, $D/Dt = \partial/\partial t + U \cdot \nabla $. Then equation (\ref{momentumnewhydro})
can be rewritten,
\begin{align}
\label{momentumnewhydro1} \rho \frac{D U}{D t} + \left(\frac{\partial  \rho}{\partial t} + \nabla \cdot [ \rho U] \right) U
+ \nabla \cdot \left[  p \mathbf{I}+ \Pi  \right] \underline{-  \nabla \cdot \left[\kappa \nabla \left(\rho U \right) \right]}= 0 ,
\end{align}
which becomes, after introducing the mass-density equation (\ref{massnewhydro}),
\begin{align}
\label{momentumnewhydro2} \rho \frac{D U}{D t} + \nabla \cdot \left[  p \mathbf{I}+ \Pi  \right]
+  \underline{\left( \nabla \cdot [\kappa  \nabla \rho]  \right)U
 -  \nabla \cdot \left[\kappa \nabla \left(\rho U \right) \right]} = 0 .
\end{align}
The underlined terms are again, the additional terms due to the adoption  of the Klimontovich kinetic equation.
Taking the cross product of equation (\ref{momentumnewhydro2}) with a hydrodynamic position vector $X$,
we notice quickly that the underlined terms generate,
\begin{align}
\label{momentum_violation}
X \wedge \left\{ \left( \nabla \cdot [\kappa  \nabla \rho]  \right)U
 -  \nabla \cdot \left[\kappa \nabla \left(\rho U \right) \right] \right\} ,
\end{align}
which we cannot write in local conservative form, i.e. as $\nabla \cdot [...]$.
So these terms appear as local angular momentum
production terms, as pointed out in \cite{OttingerHennigMarioPhysRevE2009} for other families of dissipative mass continuum models.
Equation (\ref{momentumnewhydro2}) is therefore said to violate local angular momentum conservation.

To analyze the second law of thermodynamics, energy equation (\ref{energyhydro}) is first re-written:
\begin{align}
\label{energyhydro1}
\nonumber
\rho \frac{D }{D t} \left[\frac{1}{2}  U^2   +   e_{in} \right]
+ \left(\frac{\partial  \rho}{\partial t} + \nabla \cdot [ \rho U] \right)\left(\frac{1}{2}  U^2   + e_{in} \right)
 + \nabla \cdot   \left[ \left(  p \mathbf{I} + \Pi \right)
\cdot U \right] + \nabla \cdot \left[\mathbf{q} \right] &  \\
\underline{-  \nabla \cdot \left[\kappa \nabla \left(\frac{1}{2}  \rho U^2   +  \rho e_{in} \right) \right]} & = 0  ,
\end{align}
which becomes, after introducing the mass conservation equation (\ref{massnewhydro}),
\begin{align}
\label{energyhydro2}
\nonumber
\rho \frac{D }{D t} \left[\frac{1}{2}  U^2   +   e_{in} \right]
+ \nabla \cdot [\kappa  \nabla \rho]   \left(\frac{1}{2}  U^2   + e_{in} \right)
 + \nabla \cdot   \left[ \left(  p \mathbf{I} + \Pi \right)
\cdot U \right] + \nabla \cdot \left[\mathbf{q} \right] &  \\
\underline{-  \nabla \cdot \left[\kappa \nabla \left(\frac{1}{2}  \rho U^2   +  \rho e_{in} \right) \right]} & = 0  .
\end{align}
Momentum equation (\ref{momentumnewhydro2}) can be used to eliminate the macroscopic kinetic energy terms; introducing
the density equation (\ref{massnewhydro}), the energy equation is finally
\begin{align}
\label{energyhydro-last2}
\nonumber
\rho \frac{D e_{in}}{D t}+ p\rho\frac{D  \rho^{-1}}{D t}
\underline{-  2  \kappa  \nabla \rho \cdot \nabla   e_{in}
 -  \kappa \rho \nabla \cdot \nabla e_{in}  + \frac{p}{\rho} \nabla \cdot [ \kappa  \nabla \rho ] }& \\
  \underline{ - \kappa \rho \nabla U : \nabla U}
   +  \Pi
: \nabla U  + \nabla \cdot \left[\mathbf{q} \right]
   & = 0   ,
\end{align}
in which `$ :$' denotes the Frobenius inner product. In classical fluid dynamics, the
specific entropy $s$ is defined by adopting the Gibbs (local equilibrium) equation:
\begin{align}
\label{gibbsClassic}
\rho  T \frac{D s}{D t} = \rho \frac{D e_{in}}{D t} + p\rho\frac{D  \rho^{-1}}{D t} .
\end{align}
Using equation (\ref{gibbsClassic}), energy equation (\ref{energyhydro-last2}) becomes
an equation for the entropy:
\begin{align}
\label{entropyhydro2}
\nonumber
\rho   \frac{D s}{D t}
-    \underline{ \frac{\kappa c_v}{T} \nabla \cdot \nabla ( \rho  T )
 + \kappa\left(R +c_v  \right) \nabla \cdot    \nabla \rho } & \\
   \underline{ - \kappa \rho \frac{1}{T} \nabla U : \nabla U }
   + \frac{1}{T} \Pi : \nabla U
     +  \frac{1}{T}\nabla \cdot \left[\mathbf{q} \right]
   & = 0   ,
\end{align}
where we have also used $e_{in}=c_v T$. Finally, with the identity,
\begin{align}
   \frac{1}{\phi} \nabla \cdot \nabla \phi  =  \frac{\nabla \phi \cdot \nabla \phi }{ \phi^2} + \nabla \cdot \left(  \frac{\nabla \phi }{\phi}\right) ,
 \end{align}
where $\phi$ is a scalar field, the entropy equation (\ref{entropyhydro2}) takes the form:
\begin{align}
\label{entropyhydro4}
\nonumber
 \rho   \frac{D s}{D t}
 \underline{ - \rho \kappa c_v \nabla \cdot \left(\frac{\nabla ( \rho  T )}{\rho T} \right)
 + \rho \kappa\left(R +c_v  \right) \nabla \cdot  \left( \frac{\nabla \rho}{\rho} \right) } - \kappa_h  \nabla \cdot \left(\frac{ \nabla T}{T} \right)
   &=&  \\
  \underline{  \frac{ \kappa \rho}{T} \nabla U : \nabla U}
  - \frac{1}{T} \Pi : \nabla U
  + \frac{ \kappa_h + \rho\kappa c_v }{T^2}\nabla T \cdot \nabla T
 \underline{  +\left\{
      \frac{2\kappa T c_v}{ T^2} \nabla  \rho  \cdot \nabla  T
 -\frac{\rho\kappa R} {\rho^2}\nabla \rho \cdot \nabla \rho \right\} }
     .
\end{align}
From this entropy equation (\ref{entropyhydro4}) we observe that the entropic heat flux is $- (\kappa_h + \rho\kappa c_v) \nabla T $,
which shows that heat conductivity in this stochastic continuum model cannot be associated systematically with the classical heat conductivity.
The terms in the curly brackets in equation (\ref{entropyhydro4}) can be either negative or positive.
This suggests that a negative temperature or decreasing entropy can occur.
According to classical thermodynamics with the Gibbs equation (\ref{gibbsClassic}),
these terms are therefore undesirable. Entropy flux is given on the left hand side of the equality in equation
(\ref{entropyhydro4}) by terms associated with the divergence operator.
The unwanted and difficult terms in both the momentum and entropy equations
are generated by the diffusive term in the density equation (or, more precisely, the spatial diffusive term in the initial kinetic equation).

However, in making this conclusion the fact is ignored that, in this Klimontovich description, the hydrodynamic position variable $X$
represents the motion of not a given molecule with a precise mass but a collection of molecules occupying a volume that may be changing.
More precisely, the velocity $U$ appearing in the set of equations (\ref{massnewhydro}) to (\ref{energyhydro}) may be attributed
to a volume velocity $U_v$ in same way as presented in the set of equations (\ref{massnewhydrovol}) to (\ref{energyhydrovol}).
Then, because conservation properties such as mass conservation are all properties related to the mass and mass velocity/current,
we may conclude that the violation of angular momentum conservation in equation, is simply an apparent inconsistency: this is because within equation (\ref{momentumnewhydro2}) and the derivative $\partial/\partial t + U \cdot \nabla $, we have been identifying erroneously the volume velocity with the mass-velocity leading to an incorrect interpretation of the violation of the conservation rule. A clear picture is given in the next appendix.

\section{Appendix: \label{AppendixB} Verification of other mechanical properties of the full volume-based set of equations}
Mechanical and thermodynamic properties to be satisfied by a hydrodynamic model as described and listed in \cite{OttingerHennigMarioPhysRevE2009} are:
\begin{itemize}
  \item galilean invariance
  \item integrability
  \item angular momentum conservation
  \item steady rigid fluid rotation
  \item center-of-mass position
  \item second law of thermodynamics
\end{itemize}

We take the set of equations (\ref{massnewhydrovolBren})-(\ref{fluxes-bingo-volume}) in which the material derivative is with respect to $U_m$.
Note the conservative form of that set of equations as:
\begin{subequations}
\begin{align}
\label{massnewhydrovolBrenC} \frac{\partial \rho }{ \partial t} + \nabla \cdot \left[ \rho U_m \right] =0 \ ,
\end{align}
\begin{align}
\label{momentumnewhydrovolbingoC}
\frac{\partial }{ \partial t} \left[ \rho U_m \right] +  \nabla \cdot \left[ \rho U_m U_m  \right]    +  \nabla \cdot  \left[p \mathbf{I} + \mathbf{\Pi_v} \right] =0,
\end{align}
\begin{align}
\label{energyhydrovolbingoC}
\frac{\partial  }{ \partial t}  \left[ \frac{1}{2} \rho U_m^2 + \rho e_{in}  \right] + \nabla \cdot \left[ \left( \frac{1}{2} \rho U_m^2 + \rho e_{in}  \right) U_m \right]    + \nabla
\cdot   \left[ \left[p \mathbf{I} + \mathbf{\Pi_v} \right] \cdot U_v\right]
   + \nabla \cdot \left(  -\kappa_h \nabla T \right) =0,
\end{align}
\end{subequations}
where for simplicity in notation we drop the prime above variables $p'$, $e'_{in}$, $T'$ and also bar above $\rho$.
Then, according to equation (\ref{massnewhydrovolBrenC}) the mass flux is given by $\rho U_m$ and it is the same as the momentum density
appearing the left-hand-side of the momentum transport equation (\ref{momentumnewhydrovolbingoC}).

\subsection{Galilean invariance}
We consider the following transformation:
\begin{align}
\label{galitransf0}
\left\{
  \begin{array}{l l}
     \hat{t}     & = t \quad \\
     \hat{X} - c t  & = X \quad \\
     \hat{U_m} - c & = U_m  \quad \\
   \end{array}
   \right. ,
\end{align}
where $c$ is a constant velocity vector. Equations  (\ref{galitransf0}) give transformed partial time and position derivatives by,
\begin{align}
\label{galitransf1}
\left\{
  \begin{array}{l l}
    \frac{\partial }{ \partial t }  & = \quad \frac{ \partial } { \partial \hat{t} } + c \cdot \frac{ \partial } { \partial \hat{X}} \\
     \frac{ \partial } { \partial X} & =  \quad \frac{ \partial } { \partial \hat{X} }  \\
        \end{array}
   \right. ,
\end{align}
which gives the transformed material derivative as:
\begin{align}
\label{material_transform}
\frac{D}{Dt} =\frac{\partial }{ \partial t } + U_m \cdot \nabla = \frac{\partial }{ \partial \hat{t} } + \hat{U_m} \cdot \nabla=  \frac{D}{ D \hat{t} };
\end{align}
so the material derivative is galilean invariant. Then, substituting this material derivative and the change of variable (\ref{galitransf0})
in the mass and momentum equations (\ref{massnewhydrovolBren}) and (\ref{momentumnewhydrovolbingo}) closed with equation (\ref{fluxes-bingo-stress}), we obtain the same form of equations. The galilean invariance of the energy equation is seen in a similar way  using its expression written in (\ref{energyhydrovolbingochec1}) and also due to the form of diffusive fluxes in equations (\ref{fluxes-bingo-stress})-(\ref{fluxes-bingo-volume}),  and also as transport coefficients are generally assumed independent of position and time variables.

\subsection{Integrability}
Integrability is concerned with the transport equation for total mass flux.
In the model of equations (\ref{massnewhydrovolBrenC})-(\ref{energyhydrovolbingoC})
the total mass flux is $\rho U_m$. It follows that the transport equation for the total mass is just the momentum
transport equation (\ref{momentumnewhydrovolbingoC}); so it as a conservative equation.

\subsection{Angular momentum}
Conservation of angular momentum is one of the most important properties.
Taking equation (\ref{momentumnewhydrovolbingo}) to a cross product with a hydrodynamic position vector $X$:
\begin{align}
\label{momentumnewhydrovol-conserv1}
X \wedge \rho \frac{D U_m }{ D t}  = - X \wedge \nabla \cdot \left[p \mathbf{I} + \mathbf{\Pi_v} \right] ,
\end{align}
which  is equivalent to
\begin{align}
\label{momentumnewhydrovol-conserv22}
\rho \frac{ D  }{ D t} \left[ X \wedge U_m \right]   = - X \wedge \nabla \cdot
\left[p \mathbf{I} + \mathbf{\Pi_v} \right].
\end{align}
Now, for a symmetrical second order tensor $\bar {\bar{\mathrm{T}}}$, the following property holds:
\begin{align}
X \wedge  \left[\nabla \cdot \bar {\bar{\mathrm{T}}} \right] = \nabla \cdot \left[ X \wedge \bar {\bar{\mathrm{T}}} \right].
\end{align}
The pressure tensor $\mathbf{\Pi_v}$ as given in equation (\ref{fluxes-bingo-stress}) is symmetrical.
So equation (\ref{momentumnewhydrovol-conserv22})
has the following final form:
\begin{align}
\label{momentumnewhydrovol-conserv3}
\rho \frac{ D }{ D t} \left[ X \wedge U_m \right]  = -  \nabla \cdot \left[ X \wedge
\left[p \mathbf{I} + \mathbf{\Pi_v} \right] \right].
\end{align}
Therefore conservation of angular momentum principle  is satisfied.
\subsection{Center-of-mass position}
In absence of external force the center-of-mass motion must be uniform.
To this end, one expects the quantity,
\begin{align}
\label{CenterofMassB}
  B = \rho X - \rho U_m t
\end{align}
to be a conserved quantity. So we write,
\begin{align}
\label{CenterofMassB0}
\frac{ \partial  B }{ \partial t} = \frac{ \partial }{ \partial t} \left[ \rho X - \rho U_m t \right]=
 X \frac{ \partial \rho  }{ \partial t}
  -  t \frac{ \partial   }{ \partial t} \left[ \rho U_m  \right]
  -   \rho U_m ,
\end{align}
in which transport equation of mass and momentum equations (\ref{massnewhydrovolBrenC}) and (\ref{momentumnewhydrovolbingoC})
can be introduced to obtain,
\begin{align}
\label{CenterofMassB1}
\frac{ \partial  B }{ \partial t} = - X  \nabla \cdot  \left[ \rho U_m \right]
                                  + t \nabla \cdot   \left[\rho U_m U_m + \left[p \mathbf{I} + \mathbf{\Pi_v} \right] \right]
                                  -   \rho U_m.
\end{align}
Equation (\ref{CenterofMassB1}) can be written,
\begin{align}
\label{CenterofMassB2}
\frac{ \partial  B }{ \partial t} = -  \nabla \cdot  \left[ \rho X U_m \right] + \rho U_m
                                  + t \nabla \cdot   \left[\rho U_m U_m + \left[p \mathbf{I} + \mathbf{\Pi_v} \right] \right]
                                  -   \rho U_m,
                                  \end{align}
or simply
\begin{align}
\label{CenterofMassB2}
\frac{ \partial  B }{ \partial t} = -  \nabla \cdot  \left[ B U_m - t \left[p \mathbf{I} + \mathbf{\Pi_v} \right] \right],
                                                                    \end{align}
which is therefore a conservative transport equation for the quantity defined in equation (\ref{CenterofMassB}).
Consequently center-of-mass position principle is satisfied.

The fundamental reason behind satisfying the above mechanical properties
(galilean invariance, integrability, angular momentum, center-of-mass position) are understood to be the fact that the mass flux velocity $U_m$ is the same
as the momentum density velocity $U_m$. So we are verifying these properties with respect to the mass velocity rather than a volume velocity.
All these demonstrations can be obviously reproduced taking the original set of volumed-based equations (\ref{massnewhydrovol}) to (\ref{energyhydrovol}).

\subsection{Second law of thermodynamics and the apparent continuum limit volume/mass diffusion model }
For the second law of thermodynamics, momentum equation (\ref{momentumnewhydrovolbingo}) is rewritten:
\begin{align}
\label{momentumnewhydrovolbingosquare}
\rho \frac{D }{ D t} \left[ \frac{1}{2} U_m^2 \right]  =  - U_m \cdot \nabla \cdot  \left[p \mathbf{I} + \mathbf{\Pi_v} \right],
\end{align}
and then used to eliminate the macroscopic kinetic energy in the energy equation \ref{energyhydrovolbingo}, which becomes
\begin{align}
\label{energyhydrovolbingochec0}
\rho \frac{D }{ D t} \left[  e_{in}  \right]  = &  U_m \cdot \nabla \cdot  \left[p \mathbf{I} + \mathbf{\Pi_v} \right] - \nabla
\cdot   \left[ \left[p \mathbf{I} + \mathbf{\Pi_v} \right] \cdot U_v\right]
   - \nabla \cdot \left( - \kappa_h \nabla T \right).
\end{align}
Splitting the shear stress into its components due to $U_m$ and that due to $\mathbf{J}_v$, respectively, $\mathbf{\Pi}_{Um}$ and $\mathbf{\Pi_{Jv}}$,
equation (\ref{energyhydrovolbingochec0}) can be rewritten:
\begin{align}
\label{energyhydrovolbingochec1}
\rho \frac{D }{ D t} \left[  e_{in}  \right]  = &   - p \mathbf{I} : \nabla U_m  - \mathbf{\Pi}_{Um} : \nabla U_m - \mathbf{\Pi_{Jv}} : \nabla U_m \\ \nonumber
& - \nabla \cdot \left[ p \mathbf{I} \cdot \mathbf{J}_v \right] - \nabla \cdot \left[ \mathbf{\Pi}_{Um} \cdot \mathbf{J}_v \right] - \nabla \cdot \left[\mathbf{\Pi_{Jv}} \cdot \mathbf{J}_v \right] \\ \nonumber
& - \nabla \cdot \left( - \kappa_h \nabla T \right),
\end{align}
where continuity equation (\ref{massnewhydrovolBren}) can be used to eliminate the first
term on the right-hand-side. Then taking the definition of specific entropy $s$ by the Gibbs local equilibrium equation (\ref{gibbsClassic}),
equation (\ref{energyhydrovolbingochec1}) becomes
\begin{align}
\label{energyhydrovolbingochec01}
\rho  T \frac{D s}{D t}   = &  - \mathbf{\Pi}_{Um} : \nabla U_m - \mathbf{\Pi_{Jv}} : \nabla U_m \\ \nonumber
& - \nabla \cdot \left[ p \mathbf{I} \cdot \mathbf{J}_v \right] - \nabla \cdot \left[ \mathbf{\Pi}_{Um} \cdot \mathbf{J}_v \right] - \nabla \cdot \left[\mathbf{\Pi_{Jv}} \cdot \mathbf{J}_v \right] \\ \nonumber
& - \nabla \cdot \left(  -\kappa_h \nabla T \right),
\end{align}
which may be rewritten as,
\begin{align}
\label{energyhydrovolbingochec11}
\rho  T \frac{D s}{D t}   = &  - \mathbf{\Pi}_v : \nabla U_m \\ \nonumber
& - \nabla \cdot \left[ p \mathbf{I} \cdot \mathbf{J}_v \right] - \nabla \cdot \left[ \mathbf{\Pi}_v \cdot \mathbf{J}_v \right] \\ \nonumber
& - \nabla \cdot \left( - \kappa_h \nabla T \right),
\end{align}
or
\begin{align}
\label{energyhydrovolbingochec12}
\rho  \frac{D s}{D t}   + \frac{1}{T} \nabla \cdot \left[ - \kappa_h \nabla T  +  RT \kappa_m
 \nabla \rho \right] = &
- \frac{1}{T}\mathbf{\Pi}_v : \nabla U_v   - \frac{1}{T} \mathbf{J}_v  \cdot \nabla \cdot  \mathbf{\Pi}_v
\end{align}
The last term on the right-hand-side of (\ref{energyhydrovolbingochec12}), as opposed to the first term, cannot be made to satisfy the positivity structure of the entropy production. Consequently, it appears that to fully satisfy positivity of the entropy production rate we must return to the full non-local-equilibrium structure of the volume-based hydrodynamic equations that involves the evolution of the additional microscopic structure evolution equation; this is the description presented in section \ref{entropyconsitency}. We conclude that, while it may be difficult to justify a full consistency of a diffusive volume/mass model with the classical local equilibrium Gibbs equation, beyond the local-equilibrium thermodynamics, there exists a volume/mass diffusion hydrodynamic model satisfying all mechanical and thermodynamic properties.

\bibliographystyle{elsart-num}

\end{document}